\documentclass[showpacs,preprintnumbers,amsmath,amssymb]{revtex4}
\usepackage{epsfig}
\usepackage{psfrag}
\usepackage{amsfonts}
\usepackage{graphicx}
\usepackage{dcolumn}
\usepackage{bm}
\usepackage{xcolor}
\usepackage{mathtools}

\begin{document}

\title{Axion-Polaritons in the Magnetic Dual Chiral Density Wave Phase of Dense QCD}

\author{E. J. Ferrer and V. de la Incera}
\affiliation{Department of Physics and Astronomy, University of Texas Rio Grande Valley, 1201 West University Dr., Edinburg, TX 78539}

\date{\today}

\begin{abstract}
We investigate the propagation of electromagnetic radiation in the magnetic
dual chiral density wave (MDCDW) phase of dense quark matter. Considering
the theory of low-energy fluctuations in this phase, we show how linearly
polarized photons reaching the MDCDW medium couple to the fluctuation
field to produce two hybridized modes of propagation that we call in analogy
with similar phenomenon in condensed matter physics axion polaritons, one
of them being gapless and the other gapped. The gapped mode's gap is proportional
to the background magnetic field and inversely proportional to the amplitude
of the inhomogeneous condensate. The generation of axion polaritons can
be traced back to the presence of the chiral anomaly in the low-energy
theory of the fluctuations. Considering the Primakoff effect in the MDCDW
medium, we argued that axion polaritons can be generated inside quark stars
bombarded by energetic photons coming from gamma-ray bursts and point out
that this mechanism could serve to explain the missing pulsar paradox in
the galaxy center.
\end{abstract}

\maketitle

\section{Introduction}
\label{sec1}

In recent years, many efforts have been dedicated to completing the temperature-density
phase map of QCD. The regions of extremely high temperature or density
are better understood thanks to the weakening of the strong coupling by
the phenomenon of asymptotic freedom. They are described by the quark-gluon
plasma (QGP) phase at high temperature and low density or by the color
superconducting color-flavor locked (CFL) phase at asymptotically large
density and low-temperature \cite{QCDreviews}. More challenging, nonetheless,
is to determine the phases in the intermediate density-temperature regions,
where lattice QCD is not applicable due to the sign problem, so one has
to rely on nonperturbative methods and effective theories.

It has long been argued that the region of intermediate density and relatively
low temperature may feature inhomogeneous phases, many of which have spatially
inhomogeneous chiral condensates favored over the homogeneous ones. Such
spatially inhomogeneous phases have been found in the large-N limit of
QCD \cite{largeNQCD,largeN}, in NJL models \cite{DCDW,NickelPRD80,PRD82-054009,PRD85-074002},
and in quarkyonic matter
\cite{q-chiralspirals,ferrer-incera-sanchez}. In all the cases, chiral
condensates with single-modulation are energetically favored over higher-dimensional
modulations. However, single-modulated phases in three spatial dimensions
are known to be unstable against thermal fluctuations, a phenomenon known
in the literature as Landau-Peierls instability
\cite{Landau-Peirls Inst}. In dense QCD models, the Landau-Peierls instability
occurs in the periodic real kink crystal phase \cite{Hidaka}; in the Dual
Chiral Density Wave (DCDW) phase \cite{Tatsumi}, and in the quarkyonic
phase \cite{Pisarski}. The instability signals the lack of long-range correlations
at any finite temperature, hence the absence of a true order parameter.
Only a quasi long-range order remains in all these cases, a situation that
resembles what happens in smectic liquid crystals
\cite{smectic liquid crystal}.

Furthermore, magnetic fields are a common feature in the scenarios where
quark matter phases under extreme conditions are realizable (neutron stars
(NS) \cite{Nucleons-B,Quark-B} and heavy-ion collisions \cite{B-HIC}).
Hence, investigating the field effects on quark matter phases has become
a hot topic of research (see \cite{Lec-Notes,EPJ_A52} and references therein),
thereby adding an extra dimension to the QCD phase map. The effect of a
magnetic field in inhomogeneous phases has been explored in quarkyonic
matter \cite{ferrer-incera-sanchez} and in the magnetic dual chiral density
wave (MDCDW) phase \cite{KlimenkoPRD82,Topological-Transport-1,Topological-Transport-2,Bo}. Among other things,
a magnetic field may decrease the symmetry of the original theory and activate
new channels of interaction, which in turn can generate additional condensates.
For instance, in the quarkyonic phase, a magnetic field is responsible
for the appearance of a new chiral spiral between the pion and magnetic
moment condensates \cite{ferrer-incera-sanchez}. Additional condensates
also emerge in the homogeneous chiral phase \cite{AMM} and in color superconductivity
\cite{ferrer-incera,ferrer-incera-2}.

In the case of the MDCDW phase, the external magnetic field decreases the
original global symmetries of the theory and contributes to critical topological
effects. First of all, it induces an asymmetry in the spectrum of the lowest
Landau level (LLL) \cite{KlimenkoPRD82}. The asymmetry, in turn, gives
rise to a topological term in the thermodynamic potential that significantly
enhances the window of inhomogeneity \cite{KlimenkoPRD82,PLB743}. Furthermore,
in the presence of an electric field with a nonzero component in the direction
of the background magnetic field, new topological effects emerge due to
the lack of invariance of the path-integral fermion measure under the local
chiral transformation. This lack of invariance gives rise to an ill-defined
Jacobian that requires a proper regularization
\cite{Topological-Transport-1,Topological-Transport-2}. The required regularization
procedure was carried out in \cite{Topological-Transport-2} by using Fujikawa's
method \cite{Fujikawa} and a representation of the fermion Jacobian in
terms of a set of complete and orthogonal eigenfunctions of the Dirac operator
that diagonalize the fermion action and ensure unitarity. Using this approach
\cite{Topological-Transport-2}, it was extracted, in a gauge-invariant
way, the regularized contribution to the effective action, which turned
out to be the chiral anomaly in the electromagnetic sector
$(\kappa /8)\theta F^{\ast}_{\mu \nu} F^{\mu \nu}$. This interaction couples
the electromagnetic strength tensor and its dual to $\theta =qz$, with
$q$ the condensate modulation and $z$ the spatial coordinate in the direction of
the modulation. This chiral-anomalous term leads to anomalous electric
transport properties. The difference in the symmetry group between the
DCDW and the MDCDW phases and the topological nature of the latter make
them very different and physically distinguishable, despite both phases
having the same type of inhomogeneous condensate.

In addition, as shown in Ref.~\cite{Ferrer-Incera19}, the magnetic field allows the formation of new structures in the
generalized Ginzburg-Landau (GL) expansion of the MDCDW phase that are not present in the
DCDW case. These structures are consistent with the symmetry group that
remains after the explicit breaking of the rotational and isospin symmetries
by the external magnetic field. When considering the phonon fluctuations
about the inhomogeneous condensate, these new field-induced structural
terms, whose nonzero coefficients can be traced back to the presence of the LLL asymmetric spectrum, give rise to a linear transverse mode in the spectrum of the fluctuations.  The stiffness of the fluctuation's energy dispersion
in the direction perpendicular to the modulation, prevents the existence of the Landau-Peierls (LP)
instability in the MDCDW phase. This is in sharp contrast to the
DCDW case where the fluctuation spectrum is soft in the transverse direction and, as a consequence, the system exhibits the LP instability.

The results discussed in \cite{Ferrer-Incera19} were found assuming that
the only electromagnetic field in the system was the background magnetic
field. However, as mentioned above, if there is an electric field with a component parallel to
the background magnetic field, the free-energy will contain a chiral anomaly term. Under such circumnstances, some important questions emerge: Would this anomaly
affect the theory of the fluctuations in the MDCDW phase?  if it does,
what would be the physical consequences? The goal of the present paper
is to investigate these questions. To do that, we will add electromagnetic
waves to the mix to explore the consequences of matter-light interactions
for the low-energy theory of fluctuations in the MDCDW phase.

As shown below, in the presence of the background magnetic field, the fluctuation of the axion field, which is proportional
to the phonon, linearly couples to
the electric field of the electromagnetic wave via the chiral anomaly.
This coupling leads to two hybridized propagating modes of coupled axion
and photon fields that we call, inspired by condensed matter analogies,
axion polaritons (AP). One hybridized mode is gapped with a gap proportional
to the magnetic field and inversely proportional to the magnitude of the
inhomogeneous chiral condensate. A similar linear coupling between an axion
field and a photon in the presence of a magnetic field has been found in
topological magnetic insulators. There, the time-reversal symmetry is spontaneously
broken by an antiferromagnetic order, and the magnetic fluctuations couple
to an axion field that depends on the band structure
\cite{Polariton,Axionic-Polariton}.

On the other hand, the MDCDW phase exhibits three characteristics that
are significant for the astrophysics of NS: First, the critical temperature
needed to evaporate the inhomogeneous condensate for fields
$\sim 10^{18}$ G is beyond the stellar temperatures for the whole range
of densities characteristic of relatively old magnetars \cite{Gyory-Incera}; second,
it was proved in \cite{InhStars} that the maximum stellar mass of a hybrid
star with a quark-matter core in this phase satisfies the maximum mass
observation constraints ($M \gtrsim 2M_{\odot}$)
\cite{Demorest,Antoniadis}; and third, in \cite{PRD20} it was shown that
the heat capacity of a star with MDCDW matter in its core will be well
above the lower limit expected for NS ($C_{V}\gtrsim 10^{36}(T/10^{8})$
erg/K) \cite{Cv-NS}.

Hence, the robustness of the MDCDW phase at finite temperature makes
it a viable candidate for the interior phase of magnetars. Along this line,
we shall discuss here a possible consequence for the astrophysics of magnetars based on the
conversion of energetic $\gamma $-photons into gapped axion polaritons
via the so-called Primakoff effect \cite{Primakoff}. In particular, we
will point out that the discussed polariton effect could serve as an explanation
for the so-called missing pulsar problem in the galaxy center (GC).

The paper is organized as follows: In Sec.~\ref{sec2}, we introduce the Lagrangian
density that models the phonon-photon interaction in the MDCDW medium.
In Sec.~\ref{sec3}, we study the effect of the phonon fluctuations on an electromagnetic
wave propagating into the MDCDW medium. With this goal in mind, we solve the
corresponding axion-electrodynamic equations of the neutral dense medium.
As a consequence, we find two hybridized modes, one
gapped and one gapless, that are identified with similar axion-polaritons modes that appear
in condensed matter physics in the so-called topological magnetic insulators. In Sec.~\ref{sec4}, diagonalizing the effective
Lagrangian density that includes the photon-phonon interaction in the MDCDW
medium, we obtain the axion-polariton eigenfields which are complex pseudo-scalar
field. Furthermore, new conservation laws for the
axion-polariton particle number and the corresponding Noether four-current
in the weak-field approximation are obtained. Using these results, we point out a possible
astrophysical application of the transmutation of photons into AP. Finally
in Sec.~\ref{sec5}, we present our concluding remarks, and in the Appendix, we discuss
the properties of the gauge fixing conditions used in the calculations.

\section{Photon-phonon interaction in the MDCDW phase}
\label{sec2}

In the absence of photons, the low-energy theory of fluctuations in the
MDCDW phase is given by the phonon Lagrangian density
\cite{Ferrer-Incera19}
%
\begin{equation}
\label{phononLagrangian}
\mathcal{L}_{\theta}=\frac{1}{2}[(\partial _{0}\theta )^{2}-v^{2}_{z}(
\partial _{z}\theta )^{2}-v_{\bot}^{2}(\partial _{\bot }\theta )^{2}],
\end{equation}
conveniently expressed in terms of the axion field $\theta =mqu(x)$ that
is proportional to the phonon fluctuation $u(x)$
\cite{Ferrer-Incera19}.

The coefficients
%
\begin{equation}
\label{Parallel-Coef}
v^{2}_{z}= a_{4,2}+m^{2} a_{6,2} +6q^{2}a_{6.4}+3qb_{5,3},
\end{equation}
and
%
\begin{equation}
\label{Transverse-Coef}
v_{\bot}^{2}= a_{4.2}+m^{2} a_{6.2}+2q^{2} a_{6.4}+qb_{5,3},
\end{equation}
represent the group velocities (squares) in the directions parallel and
transverse to the modulation respectively. They produce an anisotropy
in the low-energy theory that reflects the breaking of the rotational symmetry
in the MDCDW phase.

The dynamical parameters $m$ and $q$ in (\ref{Parallel-Coef})-(\ref{Transverse-Coef})
are solutions of the stationary conditions
%
\begin{equation}
\label{Stationary_Cond._VTEX1}
\partial \mathcal{F}/\partial m=0, \quad \partial \mathcal{F}/
\partial q=0
\end{equation}
with $ \mathcal{F}$ the free-energy of the MDCDW GL expansion
\cite{Ferrer-Incera19}
%
\begin{eqnarray}
\label{GL-Free_Energy-qdelta_VTEX1}
\mathcal{F}&=&a_{2,0}m^{2}+b_{3,1}qm^{2}+a_{4,0} m^{4}+a_{4,2}q^{2}m^{2}+b_{5,1}qm^{4}
\nonumber
\\
&+&b_{5,3}q^{3}m^{2}+a_{6,0}m^{6}+a_{6,2}q^{2}m^{4}+a_{6,4}q^{4}m^{2},
\end{eqnarray}
where the coefficients $a$ and $b$ are functions of temperature,
$T$, baryonic chemical potential, $\mu $, and magnetic field, ${B}$. They
can be found from the thermodynamic potential of the theory
\cite{Topological-Transport-2}. In our notation, the first subindex indicates
the total order of the term (power of the order parameter plus derivatives),
and the second subindex denotes the number of derivatives in that term.
An explicit calculation of the coefficients \cite{Gyory-Incera} showed
that the power series in $q$ effectively becomes an expansion in powers
of $q/2\mu $.

Using (\ref{phononLagrangian}), the low-energy spectrum for the axion fluctuation
in the absence of photons is
%
\begin{equation}
\label{spectrum}
\epsilon \simeq \sqrt{v^{2}_{z}k^{2}_{z} +v_{\bot}^{2}k_{\bot}^{2}},
\end{equation}
with $k_{\bot}^{2}=k_{x}^{2}+k_{y}^{2}$. Noticeably, it is linear in
both parallel and transverse momenta, a property that ensures the lack
of Landau-Peierls instability \cite{Ferrer-Incera19} in the MDCDW phase.
This is in sharp contrast to the DCDW case, which has the same type of
density wave condensate $\Delta e^{iqz}$, but its transverse group velocity
vanishes. This is due to the fact that in that case the $b$ coefficients
in (\ref{Transverse-Coef}) are zero, since they are only present at
$B\neq 0$ and originate from the non-trivial topology of the LLL dynamics.
We call attention that if $B=0$ the expression at the rhs of (\ref{Transverse-Coef})
reduces to the stationary condition for $q$ in (\ref{Stationary_Cond._VTEX1})
\cite{Ferrer-Incera19}.

Let's consider now the propagation of light in the MDCDW medium. As shown
in \cite{Topological-Transport-1,Topological-Transport-2}, this situation
triggers the appearance of the chiral anomaly
%
\begin{equation}
\label{background_Chiral_anomaly_VTEX1}
\mathcal{L}_{\bar{\theta}}= \frac{\kappa}{8} \bar{\theta} F_{\mu \nu}
\tilde{F}^{\mu \nu}
\end{equation}
in the fermion effective action with $\bar{\theta}=mqz$ a background axion
field linearly proportional to the chiral condensate parameters $m$ and
$q$.

Then, when light-matter interactions are present in this phase, the low-energy
theory of the fluctuations in the MDCDW medium takes the form
%
\begin{equation}
\label{phonon-photon-Lagrangian}
\mathcal{L}_{\theta -A}=\mathcal{L}_{\theta}+ +\mathcal{L}_{A}+
\frac{\kappa}{8} \bar{\theta} F_{\mu \nu}\tilde{F}^{\mu \nu}+
\frac{\kappa}{8} \theta (x) F_{\mu \nu}\tilde{F}^{\mu \nu},
\end{equation}
where
%
\begin{equation}
\label{photonLagrangian}
\mathcal{L}_{A}=\-\frac{1}{4}F_{\mu \nu}F^{\mu \nu}+ J^{\mu }A_{\mu }%
\end{equation}
is the conventional electromagnetic action, with $J^{\mu}$ the (non-anomalous)
electromagnetic four-current found after integrating out the fermions in
the original MDCDW effective action \cite{Topological-Transport-2}. The
phonon fluctuation term ($\frac{\kappa}{8} \theta (x) F_{\mu \nu}
\tilde{F}^{\mu \nu}$) is introduced in the chiral anomaly through the shift
$z \to z+u(x)$ in $\bar{\theta}$. The coupling constant
$\kappa \equiv \frac{2\alpha}{\pi m}$ characterizes the interaction between
the axion and the photon.

\section{Axion polaritons in the MDCDW medium}
\label{sec3}

Let us explore the consequences of the photon-axion interactions. With
that aim in mind, we assume that a linearly polarized electromagnetic wave
with electric field parallel to the background magnetic field propagates
in the MDCDW medium. Then, the corresponding field equations for the axion
and electromagnetic fields are
%
\begin{equation}
\boldsymbol{\nabla} \cdot \mathbf{E}=J^{0}+\frac{\kappa}{2}\nabla
\bar{\theta} \cdot \mathbf{B}+\frac{\kappa}{2}\nabla \theta \cdot
\mathbf{B},
\label{1}
\end{equation}
\begin{equation}
\nabla \times \mathbf{B}-\partial \mathbf{E}/\partial t=\mathbf{J}-
\frac{\kappa}{2} (\frac{\partial \theta}{\partial t} \mathbf{B}+
\nabla \theta \times \mathbf{E}),
\label{2}
\end{equation}
\begin{equation}
\boldsymbol{\nabla} \cdot \mathbf{B}=0, \quad \nabla \times \mathbf{E}+
\partial \mathbf{B}/\partial t=0
\label{3}
\end{equation}
\begin{equation}
\partial _{0}^{2} \theta - v_{z}^{2} \partial _{z}^{2}\theta -v_{
\bot }^{2} \partial _{\bot}^{2}\theta +\frac{\kappa}{2} \mathbf{B}
\cdot \mathbf{E}= 0
\label{eq13}
\end{equation}

For application to NS, we should consider a neutral medium; hence we assume
that $J^{0}$ contains an electron background charge that ensures overall
neutrality
%
\begin{equation}
\label{EMterms}
J^{0}+\frac{\kappa}{2}\nabla \bar{\theta} \cdot \mathbf{B}+
\frac{\kappa}{2}\nabla \theta \cdot \mathbf{B}=0.
\end{equation}

In the presence of a static and uniform background magnetic field
$\mathbf{B}_{0}$, the coupling between the axion fluctuation and the photon
is linear, so that the linearized field equations take the form
%
\begin{equation}
\label{Wave-Eq}
\frac{\partial ^{2}\mathbf{E}}{\partial t^{2}}=\boldsymbol{\nabla}^{2}
\mathbf{E}+\frac{\kappa}{2}
\frac{\partial ^{2}\theta}{\partial t^{2}} \mathbf{B}_{0}
\end{equation}
\begin{equation}
\frac{\partial ^{2}\theta}{\partial t^{2}}- v_{z}^{2}
\frac{\partial ^{2}\theta}{\partial z^{2}}-v_{\bot }^{2} (
\frac{\partial ^{2}\theta}{\partial x^{2}}+
\frac{\partial ^{2}\theta}{\partial y^{2}})+\frac{\kappa}{2}
\mathbf{B}_{0} \cdot \mathbf{E}= 0.
\label{eq16}
\end{equation}

In momentum space these field equations can be written as
%
\begin{equation}
\label{P-Space-1}
\left (\omega ^{2}-p^{2} \right ) E-\left (\frac{\kappa}{2} \omega ^{2}
B_{0} \right ) \theta =0
\end{equation}
%
\begin{equation}
\label{P-Space-2}
-\left (\frac{\kappa}{2}B_{0} \right ) E+\left (\omega ^{2}-P^{2}
\right )\theta =0
\end{equation}
where
%
\begin{equation}
\label{P}
P^{2}=v_{z}^{2}p_{z}^{2}+v_{\bot}^{2}p_{\bot}^{2}.
\end{equation}

From (\ref{P-Space-1})-(\ref{P-Space-2}) we notice that there is a mix
between the phonon and photon modes that gives rise to two hybridized propagating
modes with eigenvalues obtained from the dispersion relation
%
\begin{equation}
\label{Dispersion0Eq}
\det \left [ {
\begin{array}{c@{\quad}c}
\omega ^{2}-p^{2} &-\kappa \omega ^{2} B_{0}/2
\\
-\kappa B_{0}/2 &\omega ^{2}-P^{2}
\end{array}
} \right ]=0
\end{equation}

The solutions of (\ref{Dispersion0Eq}) correspond to one gapless,
$\omega _{0}$, and one gapped, $\omega _{\delta}$, propagation modes, which
are respectively given by
%
\begin{equation}
\label{Frequencies}
\omega ^{2}_{0}=\omega^2 _{1}-\omega^2 _{2}, \quad \omega ^{2}_{\delta}=
\omega^2 _{1}+\omega^2 _{2}
\end{equation}
with
%
\begin{equation}
\label{A}
\omega^2 _{1}=\frac{1}{2}[p^{2}+P^{2}+(\frac{\kappa}{2} B_{0})^{2}],
\end{equation}
%
\begin{equation}
\label{B}
\omega^2 _{2}=\frac{1}{2}\sqrt{ [p^{2}+P^{2}+(\frac{\kappa}{2} B_{0})^{2}]^{2}-4p^{2}P^{2}}.
\end{equation}

The gap of the $\omega _{\delta}$ mode is field-dependent and given by
%
\begin{equation}
\label{AP-Mass}
\omega _{\delta}(\vec{p}\rightarrow 0)=\delta =\alpha B_{0}/\pi m
\end{equation}

We call these hybridized modes axion polaritons inspired by condensed matter analogies.
In condensed matter, polaritons often appear as couple modes of optical
phonons and light, magnons and light, or, as in the case of topological
magnetic insulators, as the coupled mode of light and the axionic mode
of an antiferromagnet \cite{Polaritons-CM}. It is pretty remarkable that
despite significant differences between the underlying physics of earth-bound
topological materials and the MDCDW phase of dense quark matter, a phase
that can only exist under the extreme conditions of NS interiors, their
low-energy physics is described by similar propagating modes.

It is worth noticing that the AP gap (\ref{AP-Mass}) is proportional to
the magnitude of the applied magnetic field. The gap $\delta $ is present
as long as the MDCDW phase exists, i.e., as long as $m\neq 0$. On the other
hand, the gap does not explicitly depend on the modulation $q$ or the quark
chemical potential. For a magnetic field value of $B\simeq 10^{17}$ G,
$\delta $ is in the range $[0.06, 0.5]$ MeV and $m \in [23.5, 2.8]$ MeV
for intermediate baryonic densities $\rho \sim 3 \rho _{s}$
\cite{Gyory-Incera}, with $\rho _{s}$ being the nuclear saturation density.

\section{Axion-polariton eigenfields and particle-number conservation}
\label{sec4}

Our goal now is to find the propagating eigenfields inside the MDCDW dense
medium under an incident electromagnetic wave. To find the corresponding
eigenvectors we impose in (\ref{phonon-photon-Lagrangian}) the temporal
gauge $A_{0}=0$, together with the Feynman gauge (with gauge-fixing Lagrangian
density
$\mathcal{L}_{g}=(1/2)\partial _{\mu }A^{\mu}\partial _{\nu }A^{\nu}$).
Notice that taken separately, none of them fix the gauge freedom completely,
so one needs to impose them together (See Appendix~\ref{appA} for a discussion of
this point). After fixing the gauge, we obtain in the presence of the magnetic
field the following quadratic Lagrangian density in Minkowskian momentum
space,
%
\begin{equation}
\label{phonon-photon-Matrix}
\mathcal{L}_{q}=-\frac{1}{2}[\theta (p), A_{3}(p)] \left [ {
\begin{array}{c@{\quad}c}
A & D
\\
-D & C
\end{array}
} \right ] \left [ {
\begin{array}{c}
\theta (-p)
\\
A_{3}(-p)
\end{array}
} \right ]
\end{equation}
with
%
\begin{equation}
\label{A26}
A=p_{0}^{2}-P^{2}, \quad C=p_{0}^{2}-(\vec{p})^{2}, \quad D=(2i
\kappa )B_{0}p_{0}
\end{equation}
and $P^{2}$ given in (\ref{P}).

As can be seen from (\ref{phonon-photon-Matrix}), the chiral anomaly term
produces a mixture between the phonon field and the longitudinal electromagnetic
field component (The transverse components of the electromagnetic field
do not mix with the phonon). Thus, to find the system eigenvectors we need
to rotate the fields by using the following unitary transformation
%
\begin{equation}
\label{Field-Rotation}
[\theta (p), A_{3}(p)]=[\Theta _{0} (p), \Theta _{\delta}(p)] \cdot S=[
\Theta _{0} (p), \Theta _{\delta}(p)]\cdot \left [ {
\begin{array}{c@{\quad}c}
\cos \beta & i\sin \beta
\\
i\sin \beta & \cos \beta
\end{array}
} \right ]
\end{equation}
and
%
\begin{equation}
\label{Field-Rotation-2}
\left [ {
\begin{array}{c}
\theta (-p)
\\
A_{3}(-p)
\end{array}
} \right ]=S^{-1} \cdot \left [ {
\begin{array}{c}
\Theta '_{0}(-p)
\\
\Theta '_{\delta}(-p)
\end{array}
} \right ] = =\left [ {
\begin{array}{c@{\quad}c}
\cos \beta & -i\sin \beta
\\
- i\sin \beta & \cos \beta
\end{array}
} \right ] \cdot \left [ {
\begin{array}{c}
\Theta '_{0}(-p)
\\
\Theta '_{\delta}(-p)
\end{array}
} \right ]
\end{equation}

From (\ref{Field-Rotation}) and (\ref{Field-Rotation-2}) we have that the
rotated fields (i.e. the AP fields) are given in terms of the original
fields by the linear combinations
%
\begin{equation}
\label{Linear-combinations}
\Theta _{0}(p)=\theta (p)\cos \beta -iA_{3} (p)\sin \beta
\end{equation}
%
\begin{equation}
\label{Linear-combinations-1}
\Theta _{\delta}(p)=-i\theta (p) sin \beta -A_{3} (p)\cos \beta
\end{equation}
and
%
\begin{equation}
\label{Linear-combinations-2}
\Theta '_{0}(-p)=\Theta _{0}^{\ast}(-p) \quad \Theta '_{\delta}(-p)=
\Theta _{\delta}^{\ast}(-p)\,
\end{equation}

The fields $\Theta _{\delta}$ and $\Theta _{0}$ become the system eigenvectors
when the rotation angle is given by the relations
%
\begin{equation}
\label{Theta-Angle}
\cos \beta =\frac{1}{\sqrt{2}}\left [1+
\frac{A-C}{\sqrt{(A-C)^{2}-(2D)^{2}}} \right ]^{1/2}\!\!, ~\sin
\beta =\frac{1}{\sqrt{2}}\left [1-
\frac{A-C}{\sqrt{(A-C)^{2}-(2D)^{2}}} \right ]^{1/2}\!\!.
\end{equation}

It can be checked that the rotation matrix introduced in (\ref{Field-Rotation})
with coefficients given in (\ref{Theta-Angle}) diagonalizes, through a similarity
transformation, the $2 \times 2$ matrix of the quadratic Lagrangian density
(\ref{phonon-photon-Matrix}) getting the form
%
\begin{equation}
\label{phonon-photon-Matrix-2}
\mathcal{L}_{q}=-\frac{1}{2}[\Theta _{0}(p), \Theta _{\delta}(p)].
\left [ {
\begin{array}{c@{\quad}c}
\lambda _{1} & 0
\\
0 & \lambda _{2}
\end{array}
} \right ] \left [ {
\begin{array}{c}
\Theta ^{\ast}_{0}(-p)
\\
\Theta ^{\ast}_{\delta}(-p)
\end{array}
} \right ]
\end{equation}
with eigenvalues
%
\begin{equation}
\label{lambda-1}
\lambda _{1}= \frac{1}{2}\left [(A+C)+\sqrt{(A-C)^{2}-(2D)^{2}}
\right ],
\end{equation}
%
\begin{equation}
\label{lambda-2}
\lambda _{2}= \frac{1}{2}\left [(A+C)-\sqrt{(A-C)^{2}-(2D)^{2}}
\right ]
\end{equation}
When taking the dispersion relations $\lambda _{1}=0$ and
$\lambda _{2}=0$, the modes $\omega _{0}$ and $\omega _{\delta}$ given
in (\ref{Frequencies}) are obtained, as it should be expected.

This result implies that linearly polarized waves with their electric field
along the background magnetic field propagate in the MDCDW medium via the
$\Theta _{\delta}$ and $\Theta _{0}$ AP modes. A peculiarity of the phonon-photon
mixing in this phase is that it is kinematic in the sense that the mixing
angle explicitly depends on the momenta. This is in contrast to other mixing
mechanisms, as for instance, the one driving the mixing between the
$W^{3}_{\mu}$ and $B_{\mu}$ bosons in the electroweak theory, which takes
place with a constant mixing angle (the Weinberg angle) and that gives
rise to the massive $Z_{\mu}$ boson plus the massless photon field
$A_{\mu}$.

From (\ref{Linear-combinations})-(\ref{Linear-combinations-2}) we see that
the AP fields are a mixture of a pseudoscalar field $\theta$ and the third
component of a four-vector field $A_{\mu}$. Notice, nonetheless, that each
component of the linear combinations (\ref{Linear-combinations})-(\ref{Linear-combinations-2}),
transform identically under the discrete transformations C, P, T. To treat
independently the different components of the electromagnetic field is
feasible in this case because the Lorentz symmetry is broken at finite
density, which differentiates $A_{0}$ from the rest, and the rotational
symmetry is also broken due to the uniform magnetic field along the third-spatial
direction, which makes special the longitudinal $A_{3}$ component. Similar
characteristics for the mixing between photons and axion fields have been
found in other contexts \cite{Mixing}.

In addition to the free Lagrangian density (\ref{phonon-photon-Matrix-2}),
the effective Lagrangian density of the low-energy axion fluctuations (\ref{phonon-photon-Lagrangian})
has an interaction term, which in momentum space and in terms of the AP
fields is given by
%
\begin{eqnarray}
\label{Int-Lag}
\mathcal{L}_{int}&=&\frac{i}{2}\kappa (p_{0}+p_{0}')F_{12}(k') [(-i
\cos \beta \sin \beta ) \Theta _{0}(k)\Theta _{0}^{\ast }(-k-k')\nonumber\\
&+&(
\cos ^{2}) \beta \Theta _{0}(k)\Theta ^{\ast}_{\delta }(-k-k')+(\sin ^{2} \beta ) \Theta _{\delta}(k)\Theta _{0}^{\ast}(-k-k')
\nonumber
\\
&+&(i
\cos \beta \sin \beta ) \Theta _{\delta}(k) \Theta ^{\ast}_{\delta }(-k-k')]
\end{eqnarray}

The Lagrangian density
$\mathcal{L}_{q}+\mathcal{L}_{int}$, given in \ref{phonon-photon-Matrix-2})
and (\ref{Int-Lag}) is invariant under the continuous phase global transformation
%
\begin{equation}
\label{lglobal-transformations}
\Theta _{0} \rightarrow e^{i\alpha}\Theta _{0}, \quad \Theta _{
\delta }\rightarrow e^{i\alpha} \Theta _{\delta}.
\end{equation}
This means that the AP number is conserved, with a conserved four-current
given in the weak-field approximation by
%
\begin{eqnarray}
\label{conserved-current}
J_{\mu}^{AP}&=&i(\partial ^{\mu}\Theta _{0}^{\ast})\Theta _{0}-i(
\partial ^{\mu}\Theta _{0})\Theta _{0}^{\ast }
\nonumber
\\
&+&iv_{z}^{2}(\partial ^{\|}\Theta _{\delta}^{\ast})\Theta _{\delta }-
iv_{\bot}^{2}(\partial ^{\bot}\Theta _{\delta})\Theta _{\delta}^{
\ast }%
\end{eqnarray}

Here, we introduced the notations
$\partial ^{\|} = (\partial ^{0},\partial ^{3})$ and
$\partial ^{\bot}=(\partial ^{1}, \partial ^{2})$ for the longitudinal
and transverse partial derivatives respectively with respect to the direction
of the magnetic field. In this approximation the gapless
$\Theta _{0}$ field has a free-field current, while the current associated
to the gapped field $\Theta _{\delta}$ exhibits a breaking in the rotational
symmetry due to the presence of the magnetic field. Thus, the conserved
current of the gapped field is more responsive to the magnetic field.

It is timely to discuss here an effect that is derived from the axion-photon
vertex $\frac{\kappa}{8} \theta (x) F_{\mu \nu}\tilde{F}^{\mu \nu}$ in
(\ref{phonon-photon-Lagrangian}). In the presence of an external magnetic field
$B$ one of the two $A_{\mu}$ fields entering in the vertex can be taken
as the one associated to $B$. Then, through
this vertex an incoming photon can be transformed into an axion field and
vice versa. This is known as the Primakoff effect \cite{Primakoff}. The
Primakoff effect is a mechanism that can occur in theories with a vertex
between a scalar or a pseudoscalar field and two photons so that via this
vertex and in the presence of background electric or magnetic field, the
photon is transformed into a spin-zero field. In the context of MDCDW dense
quark matter, the Primakoff effect allows incident photons to be transformed
into AP, since the axion field in this medium is split into the two AP
eigenfields. Since the numbers of AP's are conserved, the AP will accumulate
within this high-density medium.

This result can be of interest for astrophysics, since if the interior
of magnetars host quarks in the MDCDW phase, $\gamma $-photons that penetrate
and reach the quark medium with sufficient energy can be converted into
gapped AP's that will increase the star mass. We should call attention
to the fact that although the AP mass for fields of order
$\sim 10^{17}$ G is not too large (i. e. $\sim 0.5$ MeV), extragalactic
sources of gamma ray bursts (GRB) show an isotropic distribution over the
whole sky flashing with a rate of 1000/year. The energy output of these
events is $\sim 10^{56} - 10^{59}$ MeV, with photon energies of order
$ 0.1 - 1$ MeV \cite{GRB}. Hence, each one of these events can produce
at least $10^{56} - 10^{59}$ $\gamma $-photons that could be converted
into AP's by the Primakoff effect. If we assume that about only
$10 \%$ of these photons reach the star, which is a conservative estimate
if the star is in the narrow cone of a GRB beam, then at least about
$10^{55} - 10^{58}$ of those photons can reach the star per each GRB event.
If the number of created gapped AP is enough to reach the Chandrasekhar
limit, which determines the number of AP required to induce the star collapse,
the pulsars in that region will be destroyed by the $\gamma $-ray bombardment.
In a separate publication we will discuss how this scenario can serve to
give an alternative explanation to the astronomical puzzle called the missing
pulsar problem, which refers to the failed expectation to observe a large
number of pulsars within the distance of 10 pc of the galactic center.
Theoretical predictions have indicated that there should be more than
$10^{3}$ active radio pulsars in that region \cite{G-M-P}, but these numbers
have not been observed.

\section{Concluding remarks}
\label{sec5}

This work explores the effect of the chiral anomaly in the low-energy theory
of fluctuations about the inhomogeneous ground state of the MDCDW phase
and how that affects the propagation of electromagnetic waves in that medium.
Two main results came out of this investigation. First, we demonstrated
that linear electromagnetic waves entering the MDCDW medium mix up with
the axion field giving rise to two collective hybridized AP modes. Second,
we showed that the number density of AP is a conserved quantity. Hence,
the linearized photons that penetrate the MDCDW medium produce through
the Primakoff effect \cite{Primakoff} a number of AP's that will be conserved.
If the number of photons is equally split into massless and massive AP's,
a significant part of the photon energy will be converted to mass, what
can have consequences for the physics of NS under $\gamma $-ray radiation
as pointed out above.

As discussed in the paper, the fluctuation theory can be written in terms
of a pseudoscalar field $\theta $, which is on the other hand an axion
field. In the absence of photons, the low-energy Lagrangian of
$\theta $ reduces to an anisotropic but free theory. Thanks to the magnetic
field, the dispersion modes of the axion field are stiff enough in the
direction transverse to the modulation to prevent the destabilization of
the ground state by thermal fluctuations at low temperatures. This property
ensured the absence of the Landau-Peierls instabilities that usually affect
single-modulated condensates, a fact that underlines the robustness of
the MDCDW inhomogeneous condensate at low temperatures
\cite{Ferrer-Incera19}.

Things become even more interesting when the MDCDW medium interacts with
photons. When the medium interacts with linear electromagnetic waves whose
electric field is parallel to the background magnetic field, the chiral
anomaly is activated, and as a consequence, the Lagrangian of the fluctuations
acquires a new term that couples the wave's electric field with
$\theta $. Such an anomaly-induced coupling gives rise to coupled equations
of motions between the photon and the phonon, which once diagonalized give
energy dispersions for two hybridized collective modes called AP, one gapped
and one gapless, represented respectively by the complex-conjugate eigebfields,
$\Theta _{\delta}$ and $\Theta _{0}$.

It is worth comparing the effects produced by the magnetic field and the
chemical potential on homogenous and inhomogeneous chiral condensates.
In a massless theory of fermions, an external magnetic field strengthens
the condensate because the dimensional reduction that characterizes the
dynamics of the LLL fermions implies that their lowest energy state has
no separation from the energy of the LLL antiparticles in the Dirac sea,
so pairing among them is enhanced forming a chiral condensate even at the
subcritical coupling. This is the well-known phenomenon of magnetic catalysis
of chiral symmetry breaking (MC$\chi $SB) \cite{MC}, which takes place
in any relativistic theory of interactive massless fermions in a magnetic
field. On the other hand, a chemical potential $\mu $ increases the energy
separation between particles and antiparticles, hence stressing the pairing
up to the point that the homogeneous condensate is not energetically favored
any longer. The picture gets more complicated when one considers the effect
of the magnetic field on the coupling constant. In this case, the chiral
condensate decreases with the magnetic field, a phenomenon known as inverse
magnetic catalysis \cite{IMC}.

Spatially inhomogeneous chiral condensates typically emerge only when the
chemical potential reaches a specific value, at which a Fermi surface is
formed, and the inhomogeneous condensate is favored over the restored phase.
In the case of the MDCDW condensate, however, the inhomogeneity is present
even in the region of smaller chemical potentials because a small anomalous fermion
number is generated by the anomalous term in the thermodynamic potential that is a direct consequence of the spectral
asymmetry of the LLL. This result is so robust that it exists at both overcritical
\cite{KlimenkoPRD82} and subcritical coupling \cite{Bo}. When the chemical
potential reaches certain value in the intermediate region where a homogeneous condensate
would have normally been erased, the modulation of the inhomogeneous condensate
of the MDCDW jumps to a larger value and continue increasing with $\mu$. In this
region, there is a competition between the particle-antiparticle pairing
that tends to decrease the magnitude of the condensate and the particle-hole
pairing that tends to increase its modulation \cite{Gyory-Incera}. The jump in the modulation indicates the formation of a Fermi surface and the triggering of the particle-hole pairing. The
magnetic field, in turn, enhances the window of inhomogeneity
\cite{KlimenkoPRD82} as compared to the case without magnetic field
\cite{DCDW}. These results were all found assuming a fixed coupling constant.
It remains an open question, however, what would happen in the MDCDW phase
if one considers the magnetic field's effect on the strong coupling constant.

A relevant question that naturally emerges is the compatibility of this
quark matter phase with known NS astrophysical observations. In this sense, 
it is quite encouraging to know that the MDCDW is energetically favored over
the symmetric ground state at all the densities of relevance for compact stars, and for magnetic fields and
temperatures compatible with NS conditions \cite{Gyory-Incera}. That is,
the critical temperature to erase the inhomogeneous condensate for inner
magnetic fields $\sim 10^{17}$ G is several orders above the characteristic
temperatures of old magnetars in the whole range of expected intermediate baryonic
densities. Moreover, the MDCDW phase has already been shown to be compatible
with the observed $\sim 2M_{\bigodot}$ mass of NS \cite{InhStars}. On the
other hand, long-term observations of NS temperatures during time intervals from
months to years after accretion outburst, together with continued observations
on timescales of years, have shed some light on the limit value of the
heat capacity of NS cores. Considering these observations, it has been argued \cite{Cv-NS} that the lower-limit value ($C_{V}\gtrsim 10^{36}(T/10^{8})$
erg/K) puts out of the game matter structures that exhibit superfluidity
or superconductivity of any kind, since as showed in \cite{PRD20} all these
cases are exponentially damped. This will lead to the striking conclusion
that, if the only quark matter state to be realized in NS interior is the
color superconducting CFL/MCFL phases, quarks will be ruled out from being
the matter structure of NS cores \cite{Cv-NS}. Nevertheless, it has been recently proved \cite{PRD20} that if the NS core is formed by quarks in the MDCDW phase the heat
capacity will be well above the lower limit expected for NS
\cite{Cv-NS}. Still, other astrophysical observations, like tidal deformations
and any new, more precise determination of NS mass/radius ratios
\cite{NSradiusobservations} have yet to be considered to confirm a full compatibility of the MDCDW phase with all the relevant observations.

 \appendix
\section{Annotation on gauge fixing sufficiency}
\label{appA}

To understand and clarify the necessity to fix the Feynman gauge together
with the temporal gauge, we summarize as follows some known results from
Quantum Gauge Field Theory.

As known, the Maxwell Lagrangian density in momentum space is given by
%
\begin{equation}
\label{Maxwell-Lagrangian}
\mathcal{L}_{A}=\frac{1}{2}A^{\mu }(\partial ^{2} g_{\mu \nu} -
\partial _{\mu }\partial _{\nu})A^{\nu }%
\end{equation}

Let's start by imposing a general covariant gauge given by the Lagrangian
density
%
\begin{equation}
\label{Gauge-Lagrangian}
\mathcal{L}_{g}=-(1/2\xi )\partial _{\mu }A^{\mu }\partial _{\nu }A^{
\nu }%
\end{equation}
where $\xi $ is the gauge fixing parameter.

Adding (\ref{Gauge-Lagrangian}) to (\ref{Maxwell-Lagrangian}) we have
%
\begin{equation}
\label{Maxwell-Gauge-Lagrangian}
\mathcal{L}_{A,g}=\mathcal{L}_{A}+\mathcal{L}_{g}=\frac{1}{2}A^{\mu }[
\partial ^{2} g_{\mu \nu} -(1-\frac{1}{\xi}) \partial _{\mu }
\partial _{\nu}]A^{\nu }%
\end{equation}
Fixing the gauge parameter as $\xi =1$, we have the so-called Feynman's
gauge, and the Lagrangian density reduces to
%
\begin{equation}
\label{Maxwell-Feynmn-Lagrangian}
\mathcal{L}_{A,g_{F}}=\mathcal{L}_{A}+\mathcal{L}_{g_{F}}=\frac{1}{2}A^{
\mu }\partial ^{2} g_{\mu \nu} A^{\nu }%
\end{equation}

Now, it is easy to check that the Lagrangian density (\ref{Maxwell-Feynmn-Lagrangian})
still has the gauge freedom
%
\begin{equation}
\label{Gauge-Freedom}
A^{\mu }\rightarrow A^{\prime \,\mu }=A^{\mu}+\partial ^{\mu}\alpha , \quad
\textrm{with} \quad \partial _{\mu }\partial ^{\mu }\alpha =0
\end{equation}

Clearly, this gauge freedom can be eliminated by imposing to (\ref{Maxwell-Feynmn-Lagrangian}),
an additional gauge constraint, as for instance, the temporal gauge condition,
$A_{0}=0$, to get
%
\begin{equation}
\label{Final-Lagrangian}
\mathcal{L}_{A,g_{F,T}}=\frac{1}{2}A^{i} \partial ^{2} g_{i j} A^{j},
\quad i,j=1,2,3
\end{equation}
In this way, the Lagrangian density only depends on the two transverse
physical modes.

On the other hand, the temporal gauge alone eliminates only one degree
of freedom, leaving a Hilbert space larger than the set of physical states.
Although the temporal gauge has been used since the early history of QFT
\cite{Heisenberg}, the question of how to eliminate the extra unphysical
longitudinal degree of freedom, in this case, opens the possibility to
treat the problem by introducing different constraints
\cite{Temporal-Gauge}. Another well-known problem that arises when only
considering the temporal gauge is that the photon propagator in that gauge
exhibits a pole at the zero-momentum component, $q_{0}=0$. Integrals that
present spurious singularities as this are infrared divergent and demand
special prescriptions to provide consistent calculation procedures
\cite{Propagator-Temporal-Gauge}.

Acknowledgments: This work was supported in part by NSF grant PHY-2013222.


\begin{thebibliography}{00}

\bibitem{QCDreviews} M. Buballa, Phys.  Rep. \textbf{407} (2005) 205. K.  Fukushima and T. Hatsuda, Rept. Prog. Phys. {\bf74} (2011) 014001;  M. G. Alford, A. Schmitt, K. Rajagopal and T. Schafer, 
Rev. Mod. Phys. \textbf{80} (2008) 1455.

\bibitem{largeNQCD} D. V. Deryagin, D. Y. Grigoriev and V.  A. Rubakov, Int. J. Mod. Phys. A \textbf{7} (1992) 659; E. Shuster and D. T. Son, Nucl. Phys. B \textbf{573} (2000) 434; B.-Y. Park,  M. Rho, A. Wirzba and I. Zahed, Phys. Rev. D \textbf{ 62} (2000) 034015.

\bibitem{largeN}R. Rapp, E. Shuryak, and I. Zahed, Phys. Rev. D \textbf{63} (2001) 034008; N. V. Gubina, K.G. Klimenko, S.G. Kurbanov, and V.Ch. Zhukovsky, Phys. Rev. D \textbf{86} (2012) 085011.

\bibitem{DCDW}E. Nakano and T. Tatsumi, Phys. Rev. D \textbf{71} (2005) 114006.

\bibitem{NickelPRD80} D. Nickel, Phys. Rev. D \textbf{80} (2009) 074025; Phys. Rev. Lett. \textbf{103} (2009) 072301.

\bibitem{PRD82-054009} S. Carignano, D. Nickel, and M. Buballa, Phys. Rev. D \textbf{82} (2010) 054009.

\bibitem{PRD85-074002} H. Abuki, D. Ishibashi, and K. Suzuki, Phys. Rev. D \textbf{85} (2012) 074002.

\bibitem{q-chiralspirals} T. Kojo, Y. Hidaka, L. McLerran and R. D. Pisarski, Nucl. Phys. A \textbf{843} (2010) 37; T. Kojo, Y. Hidaka, K. Fukushima, L. McLerran and R. D. Pisarski, Nucl. Phys. A \textbf{875} (2012) 94; T. Kojo, R. D. Pisarski and A. M. Tsvelik, Phys. Rev. D \textbf{82} (2010) 074015; T. Kojo, Nucl. Phys. A \textbf{877} (2012) 70.

\bibitem{ferrer-incera-sanchez} E. J. Ferrer, V. de la Incera and A. Sanchez, Acta Phys. Polon. Supp. \textbf{5} (2012) 679.

\bibitem{Landau-Peirls Inst} R. Peierls, Helv. Phys. Acta 7 (1934) 81;  Annales de l'Institut Henri Poincar\'e \textbf{5} (1935) 177; L. D. Landau, Phys. Z. Sowjet Union \textbf{11} (1937) 26; Zh. Eksp. Teor Fiz. \textbf{7} (1937) 627.

\bibitem{Hidaka} Y. Hidaka, K. Kamikado, T. Kanazawa, and T. Noumi, Phys. Rev. D \textbf{92}  (2015) 034003.

\bibitem{Tatsumi} T-G. Lee, E. Nakano, Y. Tsue, T. Tatsumi and B. Friman, Phys. Rev. D \textbf{92} (2015) 034024.

\bibitem{Pisarski}  R. D. Pisarski, V. V. Skokov, and A. M. Tsvelik, Phys. Rev. D \textbf{99} (2019) 074025.

\bibitem{smectic liquid crystal} R. M. Hornreich, M. Luban, and S. Shtrikman, Phys. Rev. Lett. \textbf{35}, 1678 (1975);  P. G. de Gennes and J. Prost, \emph{The Physics of Liquid Crystals}, Oxford University Press., New York, 1993.

\bibitem{Nucleons-B}	L. Dong and S. L. Shapiro ApJ. \textbf{383} (1991) 745.

\bibitem{Quark-B} E. J. Ferrer, V. de la Incera, J. P. Keith, I. Portillo, P. L. Springsteen, Phys. Rev. C \textbf{82}, (2010) 065802; L. Paulucci, E. J. Ferrer, V. de la Incera, and J. E.  Horvath, Phys. Rev. D  \textbf{83} (2011) 043009;
 E. J. Ferrer and A. Hackebill, Phys. Rev. C \textbf{99} (2019) 065803; Universe \textbf{5} (2019) 104. 

\bibitem{B-HIC}	D. McLerran and R. Venugopalan, Phys. Rev. D \textbf{49} (1994) 2233; Phys. Rev. D \textbf{49} (1994) 3352; Phys. Rev. D \textbf{50} (1994) 2225.

\bibitem{Lec-Notes}E. J. Ferrer and V. de la Incera, Lect. Notes Phys. \textbf{871} (2013) 399, arXiv:1208.5179 [nucl-th].

\bibitem{EPJ_A52}E. J. Ferrer and V. de la Incera, Eur. Phys. J. A \textbf{52} (2016) 266. 

\bibitem{KlimenkoPRD82} I. E. Frolov, V. Ch. Zhukovsky and K. G. Klimenko, Phys. Rev. D \textbf{82} (2010) 076002.

\bibitem{Topological-Transport-1} E. J. Ferrer and V. de la Incera, Phys. Lett. B \textbf{769} (2017) 208; Universe \textbf{4} (2018) 54.

\bibitem{Topological-Transport-2} E. J. Ferrer and V. de la Incera, Nucl. Phys. B \textbf{931} (2018) 192.

\bibitem{Bo} B. Feng, E. J. Ferrer, I. Portillo, Phys. Rev. D   \textbf{101} (2020) 056012.

\bibitem{AMM} E. J. Ferrer, V. de la Incera, I. Portillo and M. Quiroz Phys. Rev. D \textbf{89} (2014) 085034. 

\bibitem{ferrer-incera} B. Feng, E. J. Ferrer and V. de la Incera, Nucl. Phys. B \textbf{853} (2011) 213.

\bibitem{ferrer-incera-2}  B. Feng, E. J. Ferrer and V. de la Incera, Phys. Lett. B  \textbf{706} (2011) 232. 

\bibitem{PLB743} T. Tatsumi, K. Nishiyama and S. Karasawa, Phys. Lett. B \textbf{743} (2015) 66.

\bibitem{Fujikawa}	K. Fujikawa, Phys. Rev. Lett. \textbf{42} (1979) 1195; Phys. Rev. D \textbf{21} (1980) 2848; \textit{Path Integrals and Quantum Anomalies}. Clarendon Press, Oxford, 2004.

\bibitem{Ferrer-Incera19}E. J. Ferrer and V. de la Incera, Phys. Rev. D \textbf{102 }(2020) 014010.

\bibitem{Polariton} D. L. Mills and E. Burstein, Rep. Prog. Phys. \textbf{37} (1974) 817.

\bibitem{Axionic-Polariton} R. Li, J. Wang, X. L. Qi and S. C. Zhang, Nature Physics, \textbf{6} (2010) 284. 

 \bibitem{Gyory-Incera} W. Gyory and V de la Incera, Phys. Rev. D  \textbf{106} (2022) 016011. 
 
\bibitem{InhStars} S. Carignano, E. J. Ferrer,  V. de la Incera, and L. Paulucci, Phys. Rev. D \textbf{92} (2015) 105018.

\bibitem{Demorest} P. Demorest et al., Nature 467 (2010) 1081; Z. Arzoumanian, et al.  Astrophys. J. Suppl. 235 (2018) 37. 

\bibitem{Antoniadis} J. Antoniadis et al., Science 340 (2013) 6131.

\bibitem{PRD20} E. J. Ferrer, V. de la Incera and P. Sanson. Phys. Rev. D \textbf{103} (2021) 123013.

\bibitem{Cv-NS}  A. Cumming, E. F. Brown, F. J. Fattoyev, C. J. Horowitz, D. Page and S. Reddy, {\it Phys. Rev. C} \textbf{95} 025806 (2017).

 \bibitem{Primakoff} H. Primakoff, Phys. Rev. \textbf{81} (1951) 899.
  
 \bibitem{Polaritons-CM} R. Li, J. Wang, X.-L. Qi and S.-C. Zhang,  Nature \textbf{6} (2010) 284.
 
  \bibitem{Mixing} A. Capolupo, L. De Martino, G. Lambiase and An. Stabile, Phys. Lett. B \textbf{790} (2019) 427.

\bibitem{GRB} T. Piran, Phys. Rep. \textbf{314} (1999) 575; I. Bombaci and B. Datta,  Astropys. J. \textbf{530}  (2000) L69; P. Meszaros, Rept. Prog. Phys. \textbf{69} (2006) 2259; S. Razzaque, Journal of Physics: Conference Series \textbf{60} (2007) 111.

\bibitem{G-M-P} E. Pfahl and A. Loeb, Astropys. J.  \textbf{615} (2004) 253; J. -P. Macquart, N. Kanekar, D. Frail and S.Ransom, Astropys. J.  \textbf{715} (2010) 939; J. Dexter and R. M. O'Leary, Astropys. J.  \textbf{783} (2014) L7.  

\bibitem{MC}K. G. Klimenko, Teor. Mat. Fiz. \textbf{90} (1992) 3; V. P. Gusynin, V. A. Miransky and I. A. Shovkovy, Phys. Rev. Lett. \textbf{73} (1994) 3499; Nucl. Phys. B \textbf{563} (1999) 361; E. J. Ferrer and V. de la Incera, Phys. Lett. B \textbf{481} 481 (2000) 287; Yu. I. Shilnov, and V. V. Chitov, Phys. Atom. Nucl. \textbf{64} (2001) 2051 [Yad. Fiz. \textbf{64} (2001) 2138]; N. Sadooghi, A. Sodeiri Jalili, Phys. Rev. D \textbf{76} (2007) 065013; E. Rojas, A. Ayala, A. Bashir, and A. Raya, Phys. Rev. D \textbf{77} (2008) 093004; A. Raya and E. Reyes. Phys. Rev. D \textbf{82} (2010) 016004; D.-S Lee, C. N. Leung and Y. J. Ng, Phys. Rev. D \textbf{55} (1997) 6504; E. Rojas, A. Ayala, A. Bashir, and A. Raya, Phys. Rev. D \textbf{77} (2008) 093004;  C. N. Leung and S.-Y. Wang, Nucl. Phys. B \textbf{747} (2006) 266; E. J. Ferrer, V. de la Incera and A. Sanchez, Nucl. Phys. \textbf{864} (2012) 469; C. N. Leung, Y. J. Ng and A. W. Ackley, Phys. Rev. D \textbf{54} (1996) 4181.; E. J. Ferrer and V. de la Incera, Phys. Rev. D \textbf{58} (1998) 065008; E. Elizalde, E. J. Ferrer, and V. de la Incera, Phys. Rev. D \textbf{68} (2003) 096004; E.~J.~Ferrer, V.~de la Incera and A. Sanchez, Phys. Rev. Lett. \textbf{107} (2011) 041602.

\bibitem{IMC}  G. S. Bali, F. Bruckmann, G. Endrodi, Z. Fodor, S. D. Katz, S. Krieg, A. Schaefer and K. K. Szabo, JHEP \textbf{02} (2012) 044;  M. D'Elia, Lect. Notes Phys. \textbf{871} (2013) 181; R. L. S. Farias, K. P. Gomes, G. I. Krein and M. B. Pinto, Phys. Rev. C  \textbf{90} (2014) 025203;  A. Ayala, M. Loewe and R. Zamora, Phys. Rev. D \textbf{91} (2015) 016002; E. J. Ferrer, V. de la Incera and X. J. Wen, Phys. Rev. D \textbf{91} (2015) 054006.

\bibitem{NSradiusobservations} B. P. Abbott et al. (LIGO Scientific and Virgo Collaborations), Phys. Rev. Lett.  \textbf{119} (2017) 161101;  B. P. Abbott et al. (LIGO Scientific and Virgo Collaborations), Phys. Rev. Lett.  \textbf{121} (2018) 161101;
B. P. Abbott et al., Phys. Rev. X   \textbf{9} (2019) 011001; T. E. Riley, A. L. Watts, S. Bogdanov, P. S. Ray, R. M. Ludlam, S. Guillot et al., Astrophys. J. Lett. \textbf{887}  (2019) L21; M. C. Miller et al., Astrophys. J. Lett. \textbf{887} (2019) L24;  S. Bogdanov, S. Guillot, P. S. Ray, M. T. Wolff, D. Chakrabarty, W. C. G. Ho et al., Astrophys. J. Lett. \textbf{887} (2019) L25; O. Lourenco, C.R H. Lenzi, M. Dutra, E. J. Ferrer V. de la Incera, L. Paulucci and J. Horvath, Phys. Rev. D \textbf{103} (2021) 103010. 

\bibitem{Heisenberg} H. Weyl, Z. Phys. \textbf{56} (1929) 330; W. Heisenberg and W. Pauli, Z. Phys. \textbf{59} (1930) 168.

\bibitem{Temporal-Gauge} J. Kogut and L. Susskind, Phys. Rev. D \textbf{10} (1974) 3468; M. Creutz, I. J. Muzinich and T. N. Tudron, Phys. Rev. D \textbf{19} (1979) 531; M. Creutz, Ann. Phys. (N. Y.) \textbf{117} (1979) 471; K. Heller, Phys. Rev. D \textbf{36} (1987) 1830; G. Leibbrandt and M. Staley, Nucl. Phys. B \textbf{461} (1996) 259.

\bibitem{Propagator-Temporal-Gauge} J. Frenkel, Phys. Lett. B \textbf{85} (1979) 63; S. Caracciolo, G. Curci and P. Menotti, Phys. Lett. B \textbf{113} (1982) 311; H. P. Dahmen, B. Scholz and F. Steiner, Phys. Lett. B \textbf{117} (1982) 339; S. C. Lim, Phys. Lett. B \textbf{149} (1984) 201;  D. Zeppenfeld, Nucl. Phys. B \textbf{247} (1984) 125; H. O. Girotti and K. D. Rothe, Phys. Rev. D \textbf{33} (1986) 3132; P. V. Landshoff, Phys. Lett. \textbf{169} (1986) 69; H. Yamagishi, Phys. Lett. B \textbf{189} (1987) 161; S. L. Nyeo, Z. Phys. C \textbf{52} (1991) 685; S. C. Lim, Phys. Rev. D \textbf{48} (1993) 2957.




\end{thebibliography}
\end{document}